\documentclass[a4paper,french]{rnti}

\usepackage[T1]{fontenc}
\usepackage[utf8]{inputenc}

\usepackage{url}

\usepackage{indentfirst}
\usepackage{graphicx}
\usepackage{enumitem}
\usepackage{amssymb}
\usepackage{multirow}

\titrecourt{goldMEDAL, un modèle de métadonnées générique pour lacs de données}

\nomcourt{E. Scholly et al.}

\titre{goldMEDAL : une nouvelle contribution à la modélisation générique des métadonnées des lacs de données}

\auteur{Etienne Scholly\affil{1}\affilsep\affil{2},
        Pegdwendé N. Sawadogo\affil{1},
        Pengfei Liu\affil{1},\\
        Javier A. Espinosa-Oviedo\affil{1}\affilsep\affil{3}
        Cécile Favre\affil{1},
        Sabine Loudcher\affil{1},
        Jérôme Darmont\affil{1},\\
        Camille Noûs\affil{4}
        }

\affiliation{
    \affil{1}Université de Lyon, Lyon 2, UR ERIC\\
          \{etienne.scholly, pegdwende.sawadogo, pengfei.liu, javier.espinosa-oviedo,\\
          cecile.favre, sabine.loudcher, jerome.darmont\}@univ-lyon2.fr\\

    \affil{2}BIAL-X\\
          
    \affil{3}LAFMIA lab\\
          
    \affil{4}Laboratoire Cogitamus\\
          camille.nous@cogitamus.fr\\
 }
 
\resume{
    Nous résumons ici un article publié en 2021 dans l'atelier international DOLAP adossé aux conférences jointes EDBT et ICDT. Nous y proposons goldMEDAL, un modèle générique de métadonnées pour les lacs de données basé sur quatre concepts et une modélisation en trois niveaux~: conceptuel, logique et physique~\citep{scholly2021coining}.
} 

\summary{
    We summarize here a paper published in 2021 in the DOLAP international workshop DOLAP associated with the EDBT and ICDT conferences. We propose goldMEDAL, a generic metadata model for data lakes based on four concepts and a three-level modeling: conceptual, logical and physical~\citep{scholly2021coining}.
}

\begin{document}
    L'essor des mégadonnées a révolutionné les pratiques d'exploitation des données et a conduit à l'émergence de nouveaux concepts. Parmi eux, les lacs de données sont 
    de vastes dépôts de données hétérogènes qui peuvent être analysés par diverses méthodes~\citep{Dixon2010}. Un lac de données efficace nécessite un système de métadonnées qui répond aux nombreux problèmes posés par le traitement de données volumineuses et hétérogènes. L'étude des modèles de métadonnées des lacs de données est un sujet de recherche très actif et fait l'objet de plusieurs propositions.

    \medbreak    \medbreak

    Parmi celles-ci, le modèle MEDAL propose de représenter les données à travers trois concepts principaux~: les {\em objets} qui correspondent à un ensemble de données homogènes, les {\em représentations} qui résultent de transformations de l'objet associé et les {\em versions} qui représentent les mises à jour d'un objet   ~\citep{sawadogo2019medal}. Cependant, le modèle MEDAL ne peut pas représenter simultanément différents niveaux de granularité des données. 

    \cite{ravat2019metadata} proposent un modèle dont la principale contribution est la notion de métadonnées de {\em zone}, qui spécifie la zone où se trouvent les données (par exemple, zone de données brutes, zone de données traitées). Toutefois, ce modèle ne prend pas non plus en charge les niveaux de granularité multiples des données.
    
    \cite{Eichler2020} introduisent le modèle HANDLE, qui utilise le concept générique d'{\em entité de données} pour représenter à la fois des fichiers de données et des parties de fichiers de données. Ceci lui permet de prendre en charge n'importe quel niveau de granularité. Chaque entité de données est associée à des étiquettes qui représentent des zones, des niveaux de granularité ou des catégorisations. Cependant, HANDLE ne prend pas en compte le versionnement des données. 

    Nous constatons ainsi que les modèles de métadonnées existants (même les plus recents) sont soit adaptés à un cas d'utilisation spécifique, soit insuffisamment génériques pour gérer différents types de lacs de données, y compris notre premier modèle, MEDAL. 
   Pour traiter ce problème de généricité, nous présentons le modèle goldMEDAL, une évolution de 
   MEDAL. Ce nouveau modèle comprend trois niveaux de modélisation: conceptuel, logique et physique.
   

\medbreak
 
\begin{figure*}[hbt]
        \centering
        \includegraphics[width=0.75\textwidth]{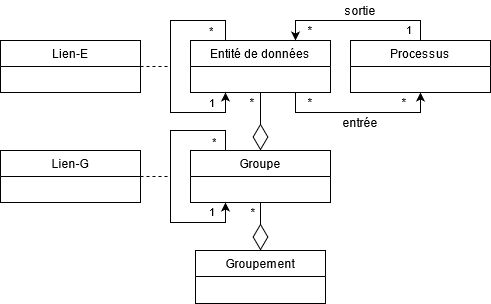} 
        \caption{Diagramme de classes UML des concepts de goldMEDAL}
        \label{fig:uml}
    \end{figure*}

    En nous inspirant des différentes notions introduites dans MEDAL, nous basons le modèle conceptuel de goldMEDAL sur quatre concepts principaux~: entité de données,  groupement, lien et processus (Figure~\ref{fig:uml}). 
    
    \begin{itemize}
        \item \textbf{Entité de données.} Les entités de données sont les unités de base du modèle de métadonnées. Elles sont flexibles en termes de granularité des données. Par exemple, une entité de données peut représenter un fichier de tableur, un document textuel ou semi-structuré, une image, une table de base de données, un tuple ou une base de données entière. L'introduction de tout nouvel élément dans le lac de données entraîne la création d'une nouvelle entité de données.
        
        \item \textbf{Groupement.} Un groupement est un ensemble de groupes ; un \textbf{groupe} rassemble des entités de données sur la base de propriétés communes. Par exemple, des zones de données brutes et prétraitées d'un lac forment les groupes d'un groupement de zones. Un autre exemple est un regroupement de documents textuels en fonction de la langue d'écriture.
        
        \item \textbf{Lien.} Les liens sont utilisés pour associer soit des entités de données entre elles, soit des groupes d'entités de données entre eux. Ils peuvent être orientés ou non. Ils permettent d'exprimer, par exemple, de simples liens de similarité entre entités de données ou des hiérarchies entre groupes. Par exemple, une hiérarchie temporelle mois $\rightarrow$ trimestre aurait les mois de janvier, février et mars liés au premier trimestre d'une année donnée.
        
        \item \textbf{Processus.} Un processus désigne toute transformation appliquée à un ensemble d'entités de données qui produit un nouvel ensemble d'entités de données.
    \end{itemize}

    \medbreak

    Au niveau logique, les concepts de goldMEDAL sont représentés à travers un graphe. 
    Ainsi, les entités de données sont représentées par des n{\oe}uds. Les liens deviennent des arêtes. Enfin, les groupes et processus sont traduits en hyper-arêtes. 
    \medbreak 
    Au niveau physique, nous avons implémenté goldMEDAL dans trois cas d'usage différents. Une première implémentation dénommée \textit{HOUDAL} et dédié à l'habitat social utilise la base de données Neo4J\footnote{\url{https://neo4j.com}} pour fournir un service de stockage et d'analyse de données principalement structurées. 
    Le lac de données \textit{AUDAL} supporte quant à lui des données tabulaires et textuelles. Il exploite les bases de données Neo4j, MongoDB\footnote{\url{https://www.mongodb.com}} et ElasticSearch\footnote{\url{https://www.elastic.co}} pour analyser l'avancée de la servicisation et la digitalisation dans les PMI de la Région Rhône-Alpes-Auvergne. 
    Enfin, la troisième implémentation, mise en \oe{}uvre au sein du projet HyperThesau et intitulée \textit{AchaeoDAL}, est dédiée à l'exploitation de données archéologiques constituées de données structurées, semi-structurées et non structurées (images et textes). Elle est basée sur Apache Atlas\footnote{\url{https://atlas.apache.org}}. 
    
    \medbreak
    \`A travers les trois modèles physiques implementés avec  goldMEDAL, nous démontrons la faisabilité ainsi que la flexibilité de notre modèle de métadonnées. De plus, les concepts de goldMEDAL généralisent ceux des modèles les plus récents~: MEDAL, \cite{ravat2019metadata} et HANDLE. Cela fait de goldMEDAL le modèle le plus générique pour la modélisation des métadonnées de lacs de données à ce jour.

\section*{Remerciements}

    Le doctorat d'E. Scholly est financé par la société  BIAL-X\footnote{\url{https://www.bial-x.com}}. Le doctorat de P.N. Sawadogo est financé par la Région Auvergne-Rhône-Alpes à travers le projet AURA-PMI. Le projet HyperThesau est financé par le Laboratoire d'Excellence “Intelligences des Mondes Urbains”\footnote{\url{https://imu.universite-lyon.fr}}.

\bibliographystyle{rnti}
\bibliography{references}


\Fr

\end{document}